# Improvements of the REDCRAFT Software Package

Casey Cole[1], Caleb Parks[1], Julian Rachele[1], Homayoun Valafar[1*]

[1]Department of Computer Science and Engineering, University of South Carolina, Columbia, SC 29208, USA

[*] **Corresponding Author Email: homayoun@cec.sc.edu Phone: 1 803 777 2404 Fax: 1 803 777 3767**
**Mailing Address: M. Bert Storey Engineering and Innovation Center, 550 Assembly St, Columbia, SC 29201**

**Abstract** – Traditional approaches to elucidation of protein structures by NMR spectroscopy rely on distance restraints also known as nuclear Overhauser effects (NOEs). The use of NOEs as the primary source of structure determination by NMR spectroscopy is time consuming and expensive. Residual Dipolar Couplings (RDCs) have become an alternate approach for structure calculation by NMR spectroscopy. In previous works, the software package REDCRAFT has been presented as a means of harnessing the information containing in RDCs for structure calculation of proteins. In this work, we present significant improvements to the REDCRAFT package including: refinement of the decimation procedure, the inclusion of graphical user interface, adoption of NEF standards, and addition of scripts for enhanced protein modeling options. The improvements to REDCRAFT have resulted in the ability to fold proteins that the previous versions were unable to fold. For instance, we report the results of folding of the protein 1A1Z in the presence of highly erroneous data.

## 1 Introduction

Faster and cheaper mechanisms of characterizing protein structures are of paramount importance in the development of personalized medicine. While there have been substantial developments in reducing the cost, and increasing the speed of sequencing genomic data, there has been little advances in improving the characterization of protein structures. In addition to the existing disparity in genetic versus proteomic information, the vast majority of the characterized protein structures belong to a very specific and limited category of proteins. For instance, while it has been estimated that 30% of the human genome encodes for membrane proteins, this important class of proteins is represented by approximately 120 proteins. Such observed disparities are rooted in the lack of new approaches to structure calculation that overcomes the existing barriers in structural determination of proteins.

In recent years, the use of Residual Dipolar Coupling (RDC) data acquired from Nuclear Magnetic Resonance (NMR) spectroscopy has become a potential avenue for a significant reduction in the cost of structure determination of proteins. In addition, RDC data have been demonstrated to overcome some standing challenges in NMR spectroscopy such as structure determination of membrane proteins, and the concurrent study of structure and dynamics of proteins. Recent work[1][2][3][4], has demonstrated the challenges in structure calculation of proteins from RDC data alone, and some potential solutions have been introduced[2][3][5]. One such approach named REDCRAFT[1][6][7][8] has been demonstrated to be successful in structure[2] calculation of proteins from a reduced set of RDC data (and therefore reduced cost). While REDCRAFT has been very successful compared to other approaches, it exhibited some limitations that resulted in reduced usability and flexibility in its use. Here we present REDCRAFT V4 that includes methodological and usability improvements. To increase the usability of REDCRAFT, we have incorporated a powerful Graphical User Interface (GUI), integrated its function with molecular visualization software, and adopted the newly approved NMR Exchange Format (NEF), to name a few. In addition to improving the usability of REDCRAFT, its core methodology has been revised to allow calculation of protein structures under challenging conditions. More specifically, we present and discuss the case of structure calculation of the protein 1A1Z using a new set of data and under lower signal to noise conditions. The REDCRAFT package is purely developed in C++ according to valid software development principles and is freely available for download via Bitbucket repository (https://bitbucket.org/hvalafar/redcraft/src/master/).

## 2 Background and Method

### 2.1 Residual Dipolar Couplings

RDCs can be acquired via NMR spectroscopy and the theoretical basis of their interaction had been established and experimentally observed in 1963[9][11]. RDC data have become a more prevalent source of data for structure determination of biological macromolecules in recent years due to the availability of alignment media and substantial improvements in NMR instruments. Upon the reintroduction of order to an isotropically tumbling molecule, RDCs can be easily acquired. The alignment medium can impose restricted tumbling through steric, electrostatic, or magnetic interaction with the protein The RDC interaction between two magnetically active nuclei can be formulated as shown in Eq. (1).

$$D_{ij} = D_{max} \langle \frac{3cos^2(\theta_{ij}(t))-1}{2} \rangle \quad (1)$$

$$D_{max} = \frac{-\mu_0 \gamma_i \gamma_j h}{(2\pi r)^3} \quad (2)$$

In this equation, $D_{ij}$ denotes the residual dipolar coupling in units of hertz between nuclei $i$ and $j$. The $\theta_{ij}$ represents the time-dependent angle of the internuclear vector between nuclei $i$ and $j$ with respect to the external magnetic field, and the angle brackets signify time averaging. In Eq. (2), $D_{max}$ represents a scalar multiplier dependent on the two interacting nuclei. In this equation, $\gamma_i$ and $\gamma_j$ are nuclear gyromagnetic ratios, $r$ is the internuclear distance (assumed fixed for directly bonded atoms), $h$ is the modified Planck's constant and $\mu_0$ represents the permeability of free space.

## 2.2 REDCRAFT Structural Fitness Calculation

While generating a protein structure from a given set of residual dipolar couplings is nontrivial, it is straightforward to determine how well a given structure fits a set of RDCs. REDCRAFT's core approach utilizes this principle in order to produce a viable protein structure. Through algebraic manipulation of Eq. (1) RDC interaction can be represented as shown in Eq. (3),

$$D_{ij} = v_{ij} * S * v_{ij}^T \quad (3)$$

where $S$ represents the Saupe order tensor matrix[9] and $v_{ij}$ denotes the normalized interacting vector between the two interacting nuclei $i$ and $j$. REDCRAFT takes advantage of this principle by quantifying the fitness of a protein to a given set of RDCs (in units of hertz) and calculating a root-mean-squared deviation as shown in Eq. (4). In this equation $D_{ij}$ and $D'_{ij}$ denote the computed and experimentally acquired RDCs respectively, $N$, represents the total number of RDCs for the entire protein, and $M$ represents the total number of alignment media in which RDC data have been acquired. In this case, a smaller fitness value indicates a better structure.

$$Fitness = \sqrt{\frac{\sum_{j=1}^{M} \sum_{i=1}^{N} (D_{ij} - D'_{ij})^2}{M*N}} \quad (4)$$

The REDCRAFT algorithm and its success in protein structure elucidation have been previously described and documented in detail [1][6][7][8][10][11][6]. Here we present a brief overview. REDCRAFT calculates structures from RDCs using two separate stages. In the first stage (*Stage-I*), a list of all possible discretized torsion angles is created for each pair of adjoining peptide planes. This list is then filtered based on allowable regions within the Ramachandran space [12][14]. The list of torsion angles that remain is then ranked based on fitness to the RDC data. These lists of potential angle configurations are used to reduce the search space for the second stage.

*Stage-II* begins by constructing the first two peptide planes of the protein. Every possible combination of angles from *Stage-I* between peptide planes $i$ and $i+1$ are evaluated for fitness with respect to the collected data, and the best $n$ candidate structures are selected, where $n$ denotes the search depth. The list of dihedral angles corresponding to the top $n$ structures is then combined with every possible set of dihedral angles connecting the next peptide plane to the current fragment. Each of these candidate structures is evaluated for fitness and the best $n$ are again selected and carried forward for additional rounds of elongation. All combination of dihedral angles worse than the best $n$ are eliminated, thus removing an exponential number of candidate structures from the search space. This elongation process is repeated iteratively, incrementally adding peptide planes until the entire protein is constructed.

## 2.3 Updates to the REDCRAFT Software Package

REDCRAFT package has been upgraded in several different categories to increase accessibility, improve usability, and enhance the core methodology. In the following sections, we highlight some examples of the improvements.

*2.3.1 Reorganization, Documentation and Addition of GUI*

The initial version of the REDCRAFT software package was only accessible through a Linux command line environment. A number of changes have been incorporated to allow REDCRAFT to be mostly platform-agnostic, and it is now compilable and executable on any Linux, BSD, or Unix system, including MacOS. Dependencies have also been updated so the project is compilable with the latest version of the GNU C Compiler. Installation of REDCRAFT now consists of two steps: generating a makefile using CMake and then running 'make install'. Utilizing CMake to generate makefiles instead of providing a static makefile allows for dynamic configurations that are suitable for each individual user's machine.

The command line environment, however, could be cumbersome to use, especially for new or novice users. To create a more streamlined analysis pipeline, the project was reorganized to allow all REDCRAFT binaries and scripts to run from a single command instead of scattered individual pieces, thereby encapsulating the project and facilitating simpler use. This is accomplished by only including a single binary, `redcraft` in the user's path that acts as command interpreter for the entire REDCRAFT project. The single redcraft binary acts as a wrapper to call any binary or script from anywhere on the computer.

Additionally, a documentation system was put in place (http://redcraft.readthedocs.io/) that allows new documentation to be built and updated upon every update to REDCRAFT. This documentation details the steps necessary to compile the entire REDCRAFT suite, as well as dependencies. The documentation may be easily exported as HTML, DOCX, or PDF document formats for offline reference.

Finally, a modern Qt5 GUI system was developed to facilitate the usage of REDCRAFT even further. The GUI, written in C++ with Qt5, is fast and available uniformly across all platforms. The GUI contains tools to run *Stage-I* and *Stage-II*, reads config files, and allows for preliminary analysis of output files. Invocation of the GUI is performed by running either `redcraft gui` or `redcraft gui [path]` (to immediately launch the GUI in that directory).

*2.3.2 Adherence to the International Standards*

The previous version of REDCRAFT utilized a rigid file format by allowing the analysis of only six classes of RDC restraints (per residue) and their corresponding uncertainties (example shown in Figure 1). These six RDC classes represented the most prevalently collected vectors in the field of NMR at the time of REDCRAFT's creation. Since then, due to advances in instrumentation, introduction of new alignment media, and data acquisition techniques, a much wider range of RDCs can be collected to aid in structure calculation. To address this issue that was common across all NMR data analysis software, the NMR community introduced the NMR Exchange Format[13] (NEF). NEF is a standard for the representation of both NMR restraints and the accompanying data. NEF was created from a series of workshops and consultations with developers of NMR structure determination software developers to streamline the pipeline of structure determination programs. The NEF formulation of RDCs is much more flexible in its definitions (an example is shown in Figure 1a). NEF lists the name, residue number, and residue name of both atoms associated with each RDC along with the RDC value and uncertainty. To accommodate the robust possibilities of RDC values that NEF could contain, REDCRAFT's computational engine was expanded to handle any combination of the interacting nuclei along the backbone of a protein. The introduction of this standard has allowed the structure determination of proteins with data that was not possible before. To remain backward compatible, a conversion script is available that will convert the legacy format into the NEF format. This conversion script has also been integrated into the GUI.

```
PHE 4.05541322826833 PHE
2 PHE C 3 PHE  N -3.03013 1
3 PHE N 3 PHE  H 0.260408 1
2 PHE C 3 PHE  H -2.49011 1
3 PHE CA 3 PHE HA -2.2099 1
3 PHE HA 3 PHE  H -6.97504 1
2 PHE HA 3 PHE  H 2.26648 1
LEU 6.18890572380312 LEU
3 LEU C 4 LEU  N -1.07715 1
4 LEU N 4 LEU  H 5.64809 1
3 LEU C 4 LEU  H -2.13885 1
4 LEU CA 4 LEU HA -12.9958 1
4 LEU HA 4 LEU  H 2.21004 1
3 LEU HA 4 LEU  H 4.7 1
```

```
PHE 4.05541322826833 PHE
-3.03013 1
0.260408 1
-2.49011 1
-2.2099 1
-6.97504 1
2.26648 1
LEU 6.18890572380312 LEU
-1.07715 1
5.64809 1
-2.13885 1
-12.9958 1
2.21004 1
4.7 1
```

(a)          (b)

*Figure 1. An example of equivalent a) NEF RDC file and b) legacy REDCRAFT RDC file.*

*2.3.3 Improvements of Decimation Methodology*

REDCRAFT's core principle approach is to generate plausible structures in a combinatorial fashion and evaluate their fitness to the experimental data. To address the intractability of combinatorial approaches, REDCRAFT has incorporated a static-decimation strategy (previously described in [1], [8]) to reduce a large number of quasi-acceptable structures into a smaller and more manageable subset of structures by selecting representative structures. The static-decimation process utilizes user-specified parameters in order to balance the two competing objectives of examining a larger pool of structures versus the computational demands of a larger and more robust search for structures. Proper selection of these parameters is normally a simple process for typical data but becomes impossible for more noisy data. Consideration of structures with poor fitness to the data is unnecessary accommodation under high signal to noise ratio. However, under the conditions of low signal-to-noise ratio, the true structure is more likely to be subjected to early elimination based on poor fitness to the data.

The new version of REDCRAFT overcomes the limitation of the static-decimation process by introducing the more intelligent dynamic-decimation process. In the dynamic decimation process, the search and decimation parameters of REDCRAFT are automatically and dynamically adjusted at each stage of the analysis to reflect the quality, and therefore the computational demands of that stage. Using the dynamic-decimation process we have investigated the low signal-to-noise instances of structure determinations that were not possible before. Here we have tested the structure determination of 1A1Z with as much as ±4 hertz of added uniform noise. This is a task that has not been successfully completed with the prior versions of REDCRAFT due to its exponentially increasing memory demands.

### 2.4 Evaluation Protocol

Our evaluation of REDCRAFT's new features and computational engine consist of three main steps. During the first step, a known protein structure is used to generate simulated RDC data. During the second step, the simulated RDC data are utilized by REDCRAFT to generate a protein structure. Finally, during the third step, the computed structure is compared to the starting structure (the ground-truth) in order to ascertain the success of REDCRAFT. Evaluation of a new methodology such as REDCRAFT based on simulate RDC data is of critical value. The use of simulated data allows for exact control over the quality of data, quantification of the performance as a function of signal-to-noise ratio, and proper assessment of time and space complexity of an algorithm as a function of data quality, to name a few.

In this project, we utilized the structure of 1A1Z to simulate RDC data in the REDCAT[14], [15] software package. Order tensors listed in Table 1 were used during the simulations. To evaluate the NEF feature of REDCRAFT the following set of RDC data {$H^\alpha$-$C^\alpha$, N-$C^\alpha$} that were unusable by the previous version of REDCRAFT were generated.

*Table 1. Order tensors used for RDC simulation.*

|  | $S_{xx}$ | $S_{yy}$ | $S_{zz}$ | $\alpha$ | $\beta$ | $\gamma$ |
|---|---|---|---|---|---|---|
| $M_1$ | $3 \times 10^{-4}$ | $5 \times 10^{-4}$ | $-8 \times 10^{-4}$ | 0° | 0° | 0° |
| $M_2$ | $-4 \times 10^{-4}$ | $-6 \times 10^{-4}$ | $10 \times 10^{-4}$ | 40° | 50° | -60° |

During the second phase of evaluation, RDC data were analyzed by REDCRAFT in order to generate the most viable protein structure. During this phase of the experiment, the REDCRAFT's RDC-fitness score was used to evaluate the success of REDCRAFT. If successful, the viable structures should exhibit an RDC-fitness to the data that is in the same order of the experimental error (related to the signal-to-noise).

During the final stage of the evaluation, the software package PyMOL[16][21] was utilized to calculate the bb-rmsd (backbone root mean squared deviation) between resulting REDCRAFT structures and the target structure. The measure of bb-rmsd is prevalently used to establish the structural similarity between two proteins and values under 3.5 angstroms can signify the success of REDCRAFT under noisy data conditions, while values under 2 angstroms can be interpreted as strong evidence for proper function and success.

## 3 Results and Discussion

### 3.1 Integration of Graphical User Interface

The Graphical User Interface (GUI), written in Qt5, was integrated seamlessly into the REDCRAFT package utilizing CMake. Qt5 contains CMake bindings to link all the necessary Qt dependencies, therefore the end user will notice no difference between compiling the REDCRAFT engine and the GUI itself. The GUI can be launched directly from the command line so that it may immediately open the current working directory, or it may be launched from its binary. REDCRAFT and subsequently REDCRAFT GUI runs seamlessly on all flavors of Linux as well as macOS. Dependencies for this version of REDCRAFT are the GCC G++ Compiler, OpenMP (used for parallelization of processing), Qt5 with Charts (for GUI support), and Python 3 and Perl (for auxiliary script support). Instructions for installation of all dependencies can be found in the REDCRAFT documentation (https://redcraft.readthedocs.io/en/latest/index.html).

After executing the GUI, the user will be presented with the screen shown in Figure 2. The initial screen consists of four panels. The first panel (Panel A) displays a greeting message as well as some "quick tips" to aid the user in utilization. Panel B loads the run parameters for *Stage-I* and *Stage-II*. Tabs allow for easy navigation between the two stages. Panel C shows all files present in the user's working directory, that is, the folder in which the REDCRAFT GUI was started in. This working directory can be changed via File->Open Directory at the top left of the GUI. In Panel D the output of each stage of execution is printed. For instance, if the "Execute Stage 1" button is pressed then the results of Stage-I angle creation will be shown (see figures 3a and 3b as examples). When in the "Stage 2" tab of Panel B, if the "Execute Stage 2" button is pressed then the results of Stage-II calculation will be shown in Panel D. When the "Advanced" tab is selected in the Stage 2 tab on Panel B, the panel expands to fill the entire column (as seen in Figure 3c) and additional parameters are shown. At any time during the execution of either stage, the process can be stopped by pressing the stage's respective "Stop" button (shown in red on Panel B).

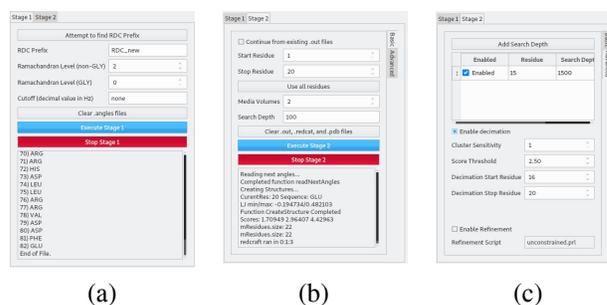

(a) (b) (c)

*Figure 3. Three examples of dialogues that can be triggered by REDCRAFT at various stages of its analysis.*

After executing the REDCRAFT analysis through its GUI, the resulting config file follows the standard INI format, but with comment support. The user is free to modify the configuration file directly, but the GUI will automatically eliminate any additional user comments in order to maintain backward compatibility.

### 3.2 Results of Structure Calculation Using Improved Decimation Method

The new version of decimation is universally faster than the previous version. Figure 4 shows the results of the first 20 residues of 1A1Z (using RDC data with ±4 hertz of error) folded with the previous version of decimation compared to the same segment folded using the new decimation method using identical search parameters. The 20-residue (out of 83 total) segment of 1A1Z was selected due to the excessive space requirement of the previous version of decimation. The previous version required 4 hours of analysis time, at the end of which the final structure exhibited a backbone RMSD of 1.589 angstroms to the reference structure (RDC fitness score of 2.21, results shown in Figure 4a). However, the extension of this fragment required memory in excess of the 16GB of the host computer and therefore did not complete the full analysis of the protein within a week. The new version of the decimation completed this exact segment on the same host computer in about 4 minutes and produced a structure with backbone RMSD similarity of 0.946 angstroms to the reference structure (RDC fitness score of 2.19, shown in Figure 4b). Of the greater importance is the success of the new version of REDCRAFT in providing a full structure of 1A1Z (illustrated in Figure 5 and discussed in the next section) that was never completed by the previous version of the software.

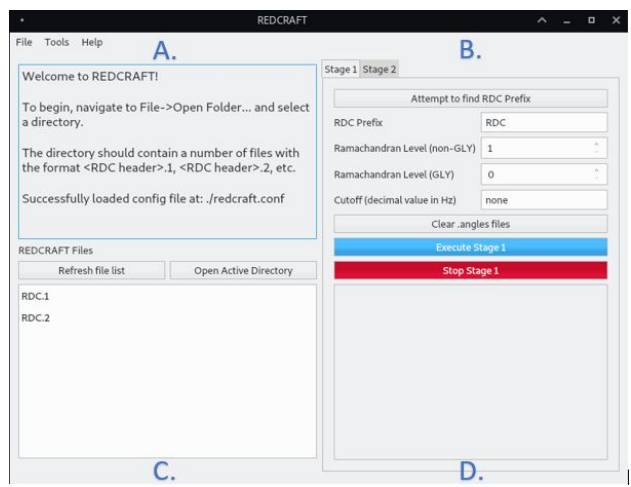

*Figure 2. The main REDCRAFT GUI implemented in Qt5.*

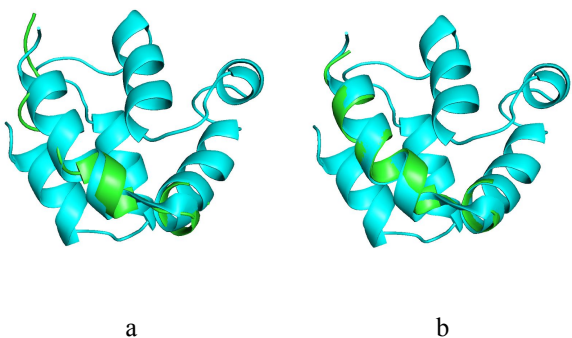

*Figure 4. Computer structure of 1A1Z with 4Hz of experimental error a) produced by the legacy version, and b) by the improved decimation procedure.*

### 3.3 Reconstruction of Proteins Using NEF Format

The changes to the core REDCRAFT engine (NEF format) enable it to perform the structure calculation of proteins based on a flexible set of RDC data. RDC pairs that were unavailable in the old version are now able to be used for reconstruction. For example, 1A1Z with {$H^\alpha$-$C^\alpha$, N-$C^\alpha$} RDC data in two alignment media with 0 hertz of simulated noise can now be folded with REDCRAFT. Using the new decimation approach, REDCRAFT produced the final structure of 1A1Z with a bb-rmsd of 1.404 angstroms and a RDC rmsd of 0.835 angstroms when compared to X-ray structure of 1A1Z (Figure 5). This is a substantial achievement in the successful folding of a protein with flexibly defined RDCs.

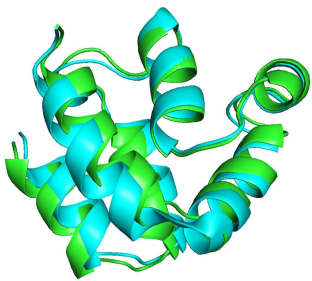

*Figure 5. A comparison of the structural similarity between the X-ray structure of 1A1Z and the computed structure of the entire structure by REDCRAFT, which has not been possible with the previous versions of REDCRAFT.*

However, it should be noted that this modification causes a slight increase in runtime that can vary from 1-5% slower than the previous version. The time requirements were benchmarked by performing structure calculation of the same protein, using the same set of RDCs in both the previous and NEF-compatible version (results shown in Figure 5). Typically, the new version of REDCRAFT completes within a minute of the previous version for an analysis that takes approximately 45 minutes, and therefore the slower performance is considered negligible.

### 3.4 Additional Scripts, Functionality, and Features

During the course of structure calculation, thousands of different phi/psi combinations are explored. Currently, the REDCRAFT algorithm will automatically generate a .pdb file for the top structure as each amino acid is added to the structure. However, one may be interested in considering an ensemble of the top *N* structures, not just the "best" structure. To facilitate this analysis, pdbgen and pdbgen2 have been added which both generate .pdb files based on a string of phi/psi angles and a string of amino acids. Pdbgen is able to generate structures directly from the .out files that are created during a run of REDCRAFT and is able to read single character residue names. Pdbgen2, which does not require any options and only takes in a string of phi/psi angles and a string of amino acids as its arguments, is simpler to use and desirable for quick pdb construction. The pdbgen collection accommodates both basic and comprehensive structure generation from phi/psi angles. These programs can also function as standalone programs for quick pdb generation and verification where the other features of REDCRAFT are not necessary. The pdbgen tools will eventually make up part of the REDCRAFT GUI analysis suite where they can be better employed to help users find exactly where the intermediate protein structure may deviate during structure generation.

### 4 Conclusions/Future Work

In this work, we have presented significant improvements to the REDCRAFT software package in the important areas of usability, accessibility, and core methodology. The inclusion of a GUI makes the software more usable by a wider audience. Incorporation of NEF standards makes the software compliant with a large suite of other widely available NMR software packages. In addition, the NEF import file allows for increased flexibility of RDCs that can be utilized by REDCRAFT which will allow structure calculations of more complex and larger proteins, such as those that have been perdeuterated due to size. We have also shown that the improved decimation method allows the method to be used to calculate proteins that it was unable to complete before due to experimental noise. Lastly, we introduced new standalone functionality to produce .pdf files from only phi/psi angles which is very useful when analyzing ensembles of structures.

## 5 Acknowledgments

This work was supported by NIH Grant Number P20 RR-016461 to Dr. Homayoun Valafar.